\documentclass[preprint,pre]{revtex4-1}

\usepackage{graphicx}

\usepackage{enumerate}
\usepackage[monochrome]{color}

\usepackage{amssymb,amsfonts,amsmath,color}

\begin{document}

\title{Noise-induced stabilization and fixation in fluctuating environment}

\author{Immanuel Meyer}
\affiliation{Department of Physics, Bar-Ilan University,
Ramat-Gan IL52900, Israel}

\author{Nadav M. Shnerb}
\affiliation{Department of Physics, Bar-Ilan University,
Ramat-Gan IL52900, Israel}


\begin{abstract}
\noindent  The dynamics of a two-species community of $N$ competing individuals is considered, with an emphasis on the role of environmental variations that affect coherently the fitness of  entire populations. The chance of fixation of a mutant (or invading)  population is calculated as a function of its mean relative fitness, the amplitude of fitness variations and their typical duration. We emphasize the distinction between the case of pairwise competition and the case of global competition; in the latter a noise-induced stabilization mechanism yields a higher chance of fixation for a single mutant. This distinction becomes dramatic in the weak selection regime, where the chance of  fixation for a single deleterious mutant is an $N$-independent constant for global competition and decays like $(\ln N)^{-1}$ in the pairwise competition case. A WKB technique yields a general formula for the chance of fixation of a deleterious mutant in the strong selection regime.  The possibility of long-term persistence of ${\cal O} (N)$ suboptimal (and extinction-prone) populations is discussed, as well as its relevance to stochastic tunneling between fitness peaks.
\end{abstract}

\maketitle

\section{Introduction}

A fundamental problem in the fields of population genetics, evolution, and community ecology, is the fate of a single mutant (or invader) introduced in a finite population of wild types. For a fixed-size community of $N$ individuals, with Markovian, zero-sum dynamics driven by stochastic birth-death events, the mutant population eventually reaches either fixation or extinction.  The classical analysis, provided by Kimura and his successors~\cite{crow1970introduction,ewens2012mathematical}, is focused on the neutral case,  (where the dynamics is only due to  demographic stochasticity, i.e., the noise  inherent in the birth-death dynamics), and on  \emph{time-independent} selective forces (deleterious/beneficial mutation).

When the system is neutral (no fitness differences, all individuals are demographically equivalent) the chance of a single mutant to reach fixation is, by symmetry, $\Pi_{n=1} = 1/N$. In general when the mutant population has abundance $n$, $\Pi_n = n/N$.

Under fixed selection $s$ the fixation probability is,
\begin{equation} \label{eq1}
\Pi_{n} = \frac{1-e^{-sn}}{1-e^{-Ns}}.
\end{equation}
In the weak selection regime, $Ns \ll 1$ and $s \ll 1$, the effect of selection  is negligible and  the neutral result reemerges. When the mutation is beneficial ($s>0$, but still $s \ll 1$) and $Ns \gg 1 $ (strong selection regime), $\Pi_{n=1}$ is $N$-independent and converges, at large $N$, to $s$. A simple and intuitive argument for this result relays on the  distinction between the region $1 \le n \le 1/s$, where demographic fluctuations are dominant and the dynamics is more or less neutral, and the region $1/s < n$, where selection dominates and fixation is almost assured. The chance of fixation is thus determined by the chance to cross a neutral region of length $1/s$, which is exactly $s$~\cite{desai2007speed}.

When the mutation is deleterious ($s<0$),  $\Pi_{n=1}$ decays exponentially with $N|s|$ in the strong selection regime, since now fixation   involves a stochastic tunneling against a constant bias. Accordingly, for any practical purpose one may neglect the chance of a deleterious mutation to reach fixation when $N$ is large. This observation poses a serious question to the standard theory of species evolution. If genotypes of  existing species are associated with  local maxima in the fitness landscape, evolutionary pathways must cross fitness valleys. Because the chance of such  tunneling events is vanishingly small, the timescales associated with it turn out to be unrealistically high \cite{gavrilets2010high}.

This set of results  was obtained for a system with pure demographic noise, where the stochastic component in the reproductive success of each individual is independent of the success of its conspecific. As a result, the per-generation noise-induced abundance variations are ${\cal O} (\sqrt{N})$. Environmental changes that affect coherently the fitness of entire populations lead to much stronger, ${\cal O} (N)$, abundance variations~\cite{lande2003stochastic}, therefore one would expect a substantial impact of these fluctuations on the chance of  fixation. Recently, many empirical studies showed that  the effect of  coherent fitness fluctuations is indeed much more pronounced than that of the demographic noise \cite{leigh2007neutral,bell2010fluctuating,kalyuzhny2014niche,kalyuzhny2014temporal,
chisholm2014temporal}. Consequently, the study  of temporal environmental stochasticity received a considerable  attention~\cite{kessler2014neutral,kessler2015neutral,saether2015concept,cvijovic2015fate,
kalyuzhny2015neutral,danino2016effect,danino2016stability,
fung2016reproducing,hidalgo2017species,danino2017environmental}. In parallel, a few recent experimental studies have considered  the response of microorganism   communities and their evolutionary pathways to fitness fluctuations~\cite{dean2017fluctuating,steinberg2016environmental,taute2014evolutionary}.

In some scenarios selection  may activate a noise-induced stabilizing mechanism that could change dramatically the chance of fixation. The main aim of this work is do provide the results for these cases and to contrast them with the known results that were obtained in the absence of such a stabilizing mechanism. As we shall see, our analysis suggest a new mechanism that allows for long-term persistence of suboptimal mutant populations, a phenomenon that may facilitate stochastic tunneling through fitness valleys (see discussion section).

To begin, let us define two zero-sum competition models, one that does not allow for noise-induced stabilization (model A) and one that admits this phenomenon (model B).

In \emph{model A} competition is pairwise and selection acts linearly.  As an example one may envisage  a population of competing animals, where a random encounter between two of them may end up in a struggle over, say, a piece of food, a mate or a territory. To model such a system we assume that these "duels" occur at a constant rate between two randomly picked individuals and in each duel the loser dies and the winner produces a single offspring. If  $x = n/N$  is the population fraction of the mutant, the chance of an interspecific "duel" is $2x(1-x)$ and the chance of the mutant individual to win such a duel is defined to be $1/2 +s/4$. Accordingly, the deterministic growth/decay of $x$ (when time is measured in generations, $N$ elementary duels in each generation) satisfies the logistic equation,
\begin{equation} \label{logistic}
\dot{x} = s x (1-x).
\end{equation}
If $s$ fluctuates in time such that its mean value is zero (a time-averaged neutral model~\cite{kalyuzhny2015neutral}) the system performs an unbiased random walk along the $z=\ln[x/(1-x)]$ axis. When the mean of $s$, $s_0$, is nonzero, the random walk is biased towards either fixation or extinction.

In \emph{model B}, on the other hand, the competition is global. In a forest, for example, following the death of an adult tree local seeds or seedlings are competing for the opened gap. If the seed dispersal length is larger than the linear size of the forest, the seed bank at any given location reflects the composition of the whole community. Death events are assumed to be random and fitness-independent. Accordingly, the chance of a species with relative log-fitness $s$ to gain one individual in an elementary death-birth event is $(1-x)x e^s/(1-x+xe^s)$, while its probability to lose one is $(1-x)x /(1-x+xe^s)$. In contrast with model A, here the fitness dependence is nonlinear.  As a result, the deterministic dynamics satisfies, to second order in $s$,
\begin{equation} \label{storage}
\dot{x} = s x (1-x) + s^2 x(1-x)(1-2x).
 \end{equation}
While the linear ($s$) term in Eq. (\ref{storage})  gives, as in Eq. (\ref{logistic}), a flow towards zero or one, the $s^2$ term has an attractive fixed point at $x=1/2$. When $s$ is fixed in time this second term is negligible, but when $s$ fluctuates the $s^2$ term may dominate. Therefore, in model B  environmental variations may induce stability through nonlinear fitness dependence~\cite{chesson1981environmental}.

While for pairwise competition (model A) the effect of environmental fluctuations weakened when their correlation time decreases, the stabilizing effect of the global competition model reaches its maximum when the correlation time is minimal~\cite{danino2016stability}.

In the next section we define mathematically these two models and explain the methodology used to obtain the chance of fixation when the environment fluctuates. The results are presented in sections \ref{resA} and \ref{resB} and the possibility of stochastic tunneling is considered  in the discussion.

\section{Model definitions and the Backward Kolmogorov Equation}

For both model A and model B  we assume that $s(t) = s_0 + \zeta(t)$, where $s_0$ is the mean log-fitness and $\zeta(t)$ may take two values, $\pm \gamma$ (telegraphic, or dichotomous, noise).

The chance of the environment to switch (from $\pm \gamma$ to $\mp \gamma$) is $1/(\delta N)$ per elementary birth-death event, so the sojourn time of the environment is taken from a geometric distribution with mean $\delta N$ elementary birth-death steps, or $\delta$ generations. The  chance of fixation when the mutant is in the plus (minus) state and its abundance is $n$, $\Pi_{+,n}$ ($\Pi_{-,n}$),  satisfies the discrete backward Kolmogorov equation (BKE),
\begin{eqnarray} \label{eq60a}
 \Pi_{+,n} &=& \left(1-\frac{1}{N \delta}\right)  \left[ W_{+,n} \Pi_{+,n+1} +W_{-,n} \Pi_{+,n+1} + (1-W_{+,n}-W_{-,n}) \Pi_{+,n} \right]  \nonumber  \\   &+&  \frac{1}{N \delta} \left[ Q_{+,n} \Pi_{-,n+1} +Q_{-,n} \Pi_{-,n+1} + (1-Q_{+,n}-Q_{-,n})  \Pi_{-,n} \right]  \nonumber \\
 \Pi_{-,n} &=& \left(1-\frac{1}{N \delta}\right) \left[ Q_{+,n} \Pi_{-,n+1} +Q_{-,n} \Pi_{-,n+1} + (1-Q_{+,n}-Q_{-,n}) \Pi_{-,n} \right]    \nonumber \\ &+&  \frac{1}{N \delta} \left[ W_{+,n} \Pi_{+,n+1} +W_{-,n} \Pi_{+,n+1} + (1-W_{+,n}-W_{-,n}) \Pi_{+,n} \right]
 \end{eqnarray}
Where $W_\pm$s are the transition probabilities in the $+\gamma$ state and the $Q_\pm$s are the probabilities in the $(-\gamma)$ states.  The transition probabilities of model A are (we replaced $n/N$ by $x$),
\begin{eqnarray}
W_{+,n}  = x(1-x)\frac{2+s_0+\gamma}{2} &\qquad& W_{-,n}  = x(1-x)\frac{2-s_0-\gamma}{2} \\ Q_{+,n} = x(1-x)\frac{2+s_0-\gamma}{2} &\qquad& Q_{-,n}  =  x(1-x)\frac{2-s_0+\gamma}{2}, \nonumber
\end{eqnarray}
and the corresponding probabilities for model B are,
\begin{eqnarray}
W_{+,n}  = \frac{ (1-x) x e^{\gamma+s_0}}{x e^{\gamma+s_0} +(1-x)} &\qquad& W_{-,n}  = \frac{ x (1-x)  }{x e^{\gamma+s_0} +(1-x)}  \\ Q_{+,n} = \frac{ (1-x) x e^{s_0}}{x e^{s_0} +(1-x)e^\gamma} &\qquad& Q_{-,n}  = \frac{ x (1-x)e^\gamma  }{x e^{s_0} +(1-x)e^\gamma}. \nonumber
\end{eqnarray}
The exact difference equations (\ref{eq60a}), with the appropriate set of $W$s and $Q$s and with the boundary conditions $\Pi_{+,0} = \Pi_{-,0} = 0$, $\Pi_{+,N} = \Pi_{-,N} = 1$, may be solved numerically as a Markov chain~\cite{ashcroft2014fixation} or as a linear system~\cite{danino2016stability}, and these solutions are compared below with the analytic formulas.

One can translate this pair of discrete BKEs for $\Pi_{\pm,n}$  into an equivalent set for $\Pi_n \equiv (\Pi_{+,n} + \Pi_{-,n})/2$ and $\Delta_n \equiv (\Pi_{+,n} - \Pi_{-,n})/2$. Taking the  continuum limit where $n$ is replaced by $Nx$ and functions of $x \pm 1/N$ are expanded to second order in $1/N$, a pair of coupled,  second order differential equations for $\Pi(x)$ and $\Delta(x)$ emerges. In~\cite{danino2016stability} we have analyzed these equations in the limit of large $N$ and small $s_0$. Using a dominant balance argument we showed that the dynamics is governed by a single second-order equation (in~\cite{danino2016stability} only the time to absorption was discussed, but the equation for $\Pi$ is the homogenous version of the corresponding BKE, see~\cite{redner2001guide}). This equation is,
\begin{equation} \label{eq10}
[1+Gx(1-x)]\Pi''(x) + \left(s_0 N + \eta G  (1-2x) \right) \Pi'(x) = 0,
\end{equation}
with the boundary conditions $ \Pi(0)=0 $  and $ \Pi(1) = 1$. Here $g \equiv \gamma^2 \delta /2$  is the strength of the environmental noise and $G \equiv Ng$ is this strength divided by the strength of demographic stochasticity $1/N$.  The differences between model A and B are encapsulated in the parameter $\eta$:
 \begin{eqnarray}
\rm{Model \ A} &\qquad&  \eta = 1 \nonumber \\
\rm{Model \ B} &\qquad& \eta =1+\frac{1}{\delta}.
 \end{eqnarray}
As $\delta$ grows, model B becomes closer to model A. However, the derivation of Eq. (\ref{eq10}) assumes that fixation cannot occur during a single sweep of the environment, so an increase in $\delta$ is legal only if $N$ increases such that $\delta \ll \ln N/(s_0+\gamma)$.

 Eq. (\ref{eq10}) may be solved using integrating factors, but this leads to complicated and hard to interpret nested integrals expressions. Instead one may analyze this equation in the inner ($x \ll 1$), middle ($Gx(1-x) \gg 1$) and outer ($1-x \ll 1$) regimes and then match asymptotically the solutions in the large $N$ (more precisely, large $G$) limit.  In the next section we present briefly the results for model A, following~\cite{danino2017fixation}. Our purpose is to contrast these result with the outcomes of model B and to emphasize the  effects of the noise-induced stabilizing mechanism.

\section{Model A: local competition and linear selection} \label{resA}

The solutions of model A in the inner, middle and outer regimes are given by,
\begin{eqnarray} \label{modelA1}
\Pi_{in}(x) \   &=& C_1 (1-(1+Gx)^{-\alpha}) \\
\Pi_{m} (x) \ &=& C_3+C_2  \left(\frac{1-x}{x}\right)^\alpha \nonumber \\
\Pi_{out}(x) &=& 1- C_4 (1-[1+G(1-x)]^{\alpha}), \nonumber
\end{eqnarray}
where $\alpha \equiv s_0/g$ and,
\begin{eqnarray} \label{modelA2}
C_1 &=& C_3 = \frac{1}{(1- G^{-2\alpha})} \nonumber \\
 C_4 &=& 1-C_1 \nonumber \\
C_2 &=& -C_1 G^{-\alpha }.
\end{eqnarray}

\begin{figure}
\includegraphics[width=8cm]{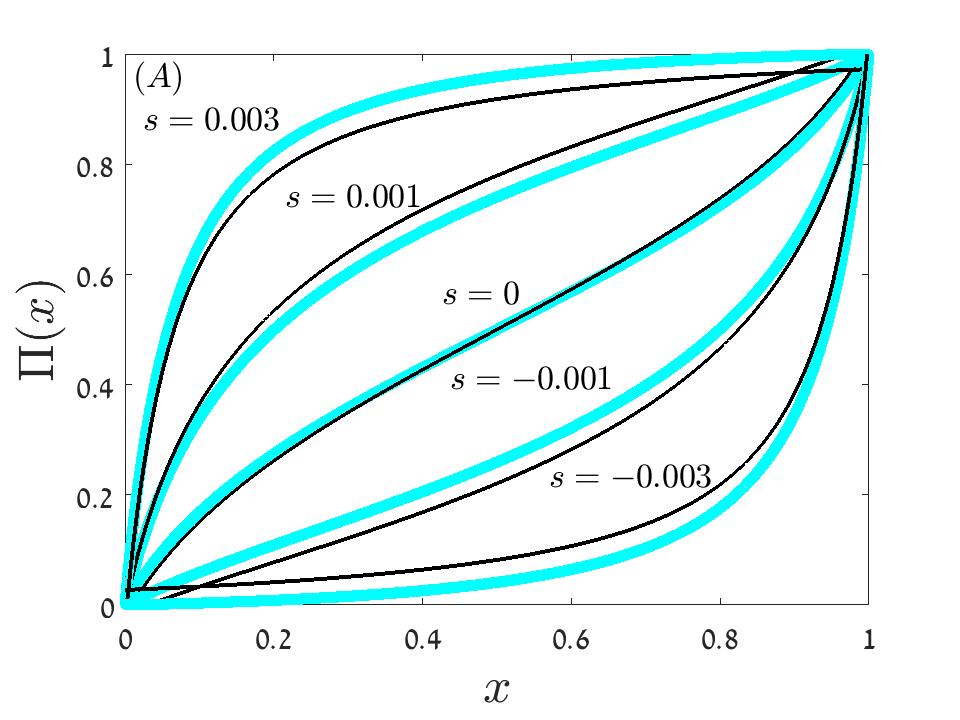}
\includegraphics[width=8cm]{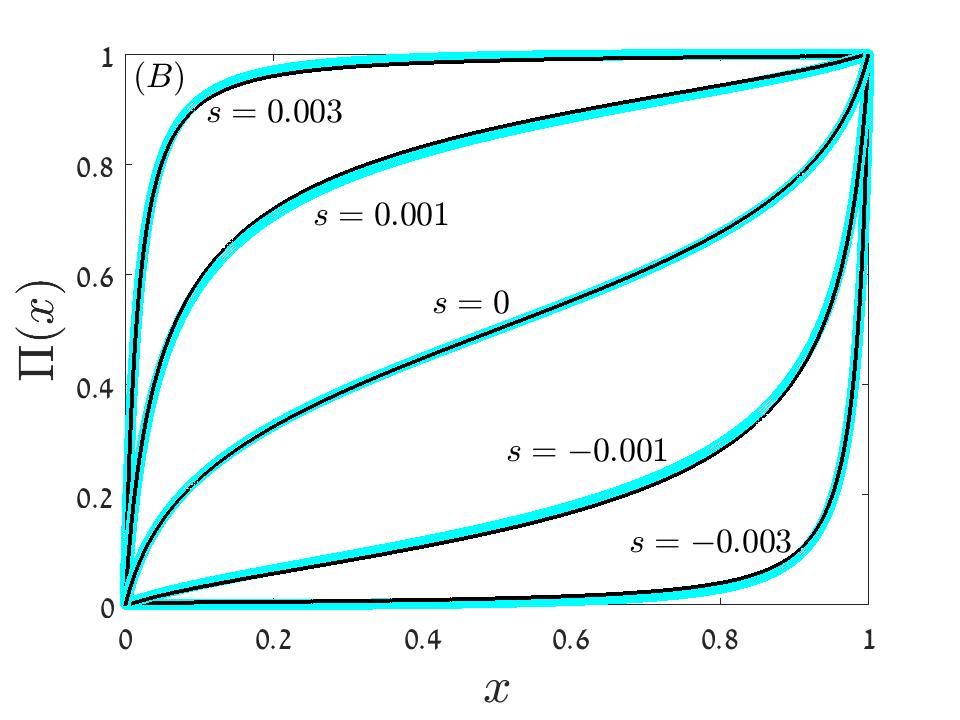}
\caption{$\Pi(x)$ vs. $x$ for model A. In both panels $\gamma = 0.2$ and $\delta = 0.1$; $N=5,000$ in panel (A)  and $N=20,000$ in panel (B). Numerical solutions of the discrete equation (\ref{eq60a}) (blue circles) are compared with the uniform approximations (\ref{unif}) and (\ref{unif1}) (black full lines) for $s=0$ ($\alpha =0$), $s= \pm 0.001$ ($\alpha=1/2$) and $s= \pm 0.003$ ($\alpha =3$). When $|\alpha| <1$ (\ref{unif}) has been used, while for $|\alpha|>1$ we implemented the uniform approximation (\ref{unif1}). For any fixed nonzero value of $s$, as $N$ grows $\Pi(x)$ sticks to either one (if $s>0$) or zero (if $s<0$) in the middle and the outer regions. The accuracy of the uniform approximation becomes better when $N$ increases. }\label{fig1}
\end{figure}

For $|\alpha| <1$ one may use the uniform approximation solution for an arbitrary $x$,
\begin{equation} \label{unif}
\Pi_{unif}(x) = C_1 \left( 1-(1+Gx)^{-\alpha}-[1+G(1-x)]^{\alpha}+
\frac{1-(x^\alpha-1)(1-x)^\alpha}{(Gx)^\alpha}\right)+[1+G(1-x)]^{\alpha}.
\end{equation}
For $|\alpha| >1$, if $C_2$ is negligible, i.e., if $G^{-|\alpha|} \ll 1$, the uniform approximation takes the form,
\begin{equation} \label{unif1}
\Pi_{unif}(x) = C_1 \left( 1-(1+Gx)^{-\alpha}-[1+G(1-x)]^{\alpha} \right)+[1+G(1-x)]^{\alpha}.
\end{equation}
The agreement between $\Pi_{unif}$ and the outcomes of the numerical solutions of Eqs. (\ref{eq60a}) is demonstrated in Figure \ref{fig1}. The theory and the numerics become closer and closer as $N$ increases.

\begin{figure}
\includegraphics[width=10cm]{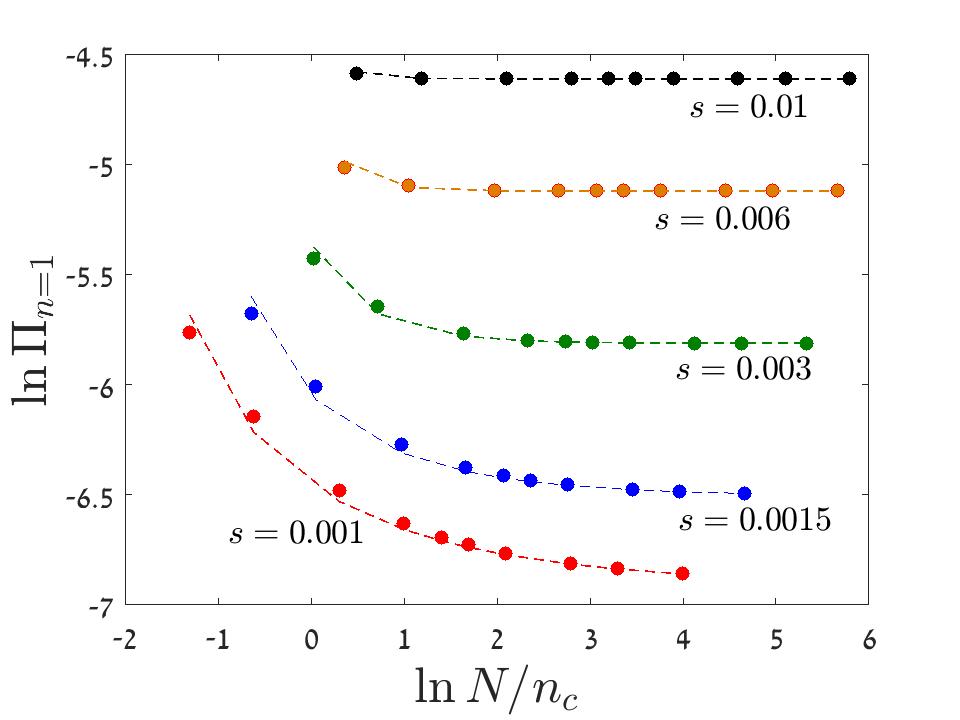}
\caption{The  chance of fixation by the lineage of a single beneficial mutant, $\Pi_{n=1}$, is plotted against the effective community size $N/n_c$ on a double logarithmic scale. Parameters are $\gamma = 0.2$, $\delta = 0.2$ and different values of $s$. Filled circles represent the results of a numerical simulation and the dashed lines are the prediction of Eq. (\ref{sm}). The actual values of $N$ used in this figure span four orders of magnitude, from $10$ to $10^5$. For $N<n_c$ the chance of fixation decays logarithmically with $N$ and $\Pi_{n=1}$  saturates to a finite value when $N>n_c$.        }\label{fig3}
\end{figure}

The chance of a single mutant to reach fixation  is obtained by plugging $x=1/N$ into the inner solution,
\begin{equation} \label{sm}
\Pi_{n=1} = \frac{1-\frac{1}{(1+g)^{s_0/g}}}{1-G^{-2\alpha}}.
\end{equation}
In figure \ref{fig3} the predictions of (\ref{sm}) are shown to fit the numerical results.

The most important moral from the comparison between  Eqs. (\ref{modelA1}, \ref{modelA2}) and Eq. (\ref{eq1}) is the modification of the criteria for strong selection. We \emph{define} the strong selection sector  as the parameter regime where  the chance of fixation of a single beneficial mutant  becomes $N$ independent. This happens when $N \gg n_c$ where $n_c$ marked the point where the deterministic effect of selection dominates against the stochastic effects of fluctuations.  While for system with selection and pure demographic noise $n_c = 1/s$, here the criteria for strong selection is $C_1 = C_3 = 1$,  i.e.,
\begin{equation} \label{nc}
n_c \sim \frac{\exp(g/2|s_0|)}{g}.
\end{equation}
This scale diverges exponentially when the mean selection is much smaller than the effective strength of fitness fluctuations, which might be the generic situation in leaving systems. Accordingly, under environmental stochasticity systems may be in the weak selection regime even if $N$ is very large.

Two other characteristic scales in this system are $N_1$ and $n_2$.  For $N \gg N_1$, $C_2 =0$, so  $N_1 \sim  \exp(g/s)/g$. The chance of fixation of the mutant population becomes large when it reaches $n_2$ such that $\Pi_{in}(n_2/N) > 1-e^{-1}$, thus,
\begin{equation} \label{nc1}
n_2 \equiv \frac{\exp^{g/s_0}-1}{g}.
\end{equation}
This scale  has been identified in~\cite{cvijovic2015fate}. Clearly, all of the three scales have similar features and they grow exponentially when $s_0$ is much smaller than $g$.  For a system with pure demographic noise and selection (Eq. \ref{eq1}), the chance of fixation becomes $N$ independent above $n_c=1/s_0$ and the condition for $n_2$ yields $1/s_0$ as well. While $n_2$ of (\ref{nc1}) converges to this limit when $g \to 0$, $n_c$ does not, and this reflect the inadequacy of our expression for $C_1$ in the $s \gg \gamma$ limit~\cite{danino2017fixation}.

In the weak selection regime, $N < n_c$, the chance of fixation is $N$-dependent. When  $\alpha \ln(G) \ll 1$  one can expand the inner solution in small $\alpha$ to obtain~\cite{cvijovic2015fate,danino2017fixation},
\begin{equation}
\Pi_{n=1} = \frac{\ln(1+g)}{2\ln G}.
\end{equation}
While $\alpha \ln G \ll 1$ is small, $\ln G$ may be large, so in the weak selection regime, and in particular in the time-averaged neutral scenario where $s_0 = 0$, the chance of fixation decays logarithmically  with system's size as demonstrated in Figure \ref{fig3}.

 When selection is weak an increase in $g$ increases the chance of fixation. In the purely demographic neutral case $\Pi$ is determined by abundance so $\Pi_{n=1} = 1/N$. In the limit of infinitely strong environmental variations a mutant will reach fixation for certainty if it was born in the right time, so the chance of fixation will grow to one half. In general the transition from abundance-dependence to environment dependence facilitates the chance of low-abundance populations to win~\cite{cvijovic2015fate}. The situation is completely different in the strong selection regime~\cite{danino2017fixation}, where $\Pi_{n=1}$ is a monotonously \emph{decreasing} function of $g$. Here the reason is the divergence of $n_c$ when $g$ increases, meaning that the chance of the beneficial mutant population to enter the region of deterministic selective growth is much smaller.

\section{Model B:  noise induced stabilization} \label{resB}

In model B the dynamics is affected by the noise-induced stabilizing mechanism that facilitates the invasion of a mutant. Before we introduce the expressions for the chance of fixation, we would like to discuss the conditions under which this stabilizing mechanism takes place.

When $s(t) = s_0 \pm \gamma$, as described above,  and the environmental fluctuations are rapid, Eq. (\ref{storage}) for the deterministic dynamics of the population takes the form
 \begin{equation}
 \dot{x} = s_0 x(1-x)+\gamma^2 x(1-x)(1/2-x).
  \end{equation}
  This equation supports an attractive fixed point  at
  \begin{equation}
 x^* = 1/2 + s_0/\gamma^2,
 \end{equation}
and  $x^*$ is between zero and one if
  \begin{equation}
  -1< \tilde{s} \equiv \frac{2 s_0}{\gamma^2} < 1.
  \end{equation}
 Therefore, $\tilde{s}$, the ratio between mean selection and  environmental fluctuations, determines the qualitative behavior of the system. When $|\tilde{s}| <1$ the noise induces a stable coexistence point and the dynamics of model B differs substantially  from the dynamics of model A. When $|\tilde{s} | > 1$ the deterministic force does not change its sign in the region between fixation and extinction, so the behaviors of model A and model B are qualitatively similar. In agreement with this observation, in~\cite{danino2016stability} the time to absorption (either fixation or extinction) for model B was found to diverge like $N^{(1-|\tilde{s}|)/\delta}$ when $N$ is large and $|\tilde{s}|<1$, while for $|\tilde{s}|>1$ the $N$ scaling is sublinear. Therefore, in this section we consider only the $|\tilde{s}|<1$ case.

Implementing  the technique of asymptotic  matching to model B equation when $|\tilde{s}|<1$, the solutions for Eq. (\ref{eq10}) with  $\eta = 1+1/\delta$ are,
\begin{eqnarray} \label{modelB1}
\Pi_{in}(x) \   &=& C_1 [1-(1+Gx)^{-\alpha-1/\delta} ]\\
\Pi_{m} (x) \ &=& C_3+C_2  \int^x \frac{(1-t)^{\alpha-\eta}}{t^{\alpha+\eta}} \ dt \nonumber \\
\Pi_{out}(x) &=& 1- C_4 (1-[1+G(1-x)]^{\alpha-1/\delta}, \nonumber
\end{eqnarray}
where as before $\alpha \equiv s_0/g$ and the constants are given by,
\begin{eqnarray} \label{modelB2}
C_1 &=& C_3 = \frac{1}{(1+ D_1 G^{-2\alpha})} \nonumber \\
 C_4 &=& 1-C_1 \nonumber \\
C_2 &=& (\alpha+1/\delta) C_1 G^{-\alpha-1/\delta },
\end{eqnarray}
with
\begin{equation}
D_1 = \frac{1/\delta+\alpha}{1/\delta-\alpha} = \frac{1+\tilde{s}}{1-\tilde{s}}.
\end{equation}

\begin{figure}
\includegraphics[width=8cm]{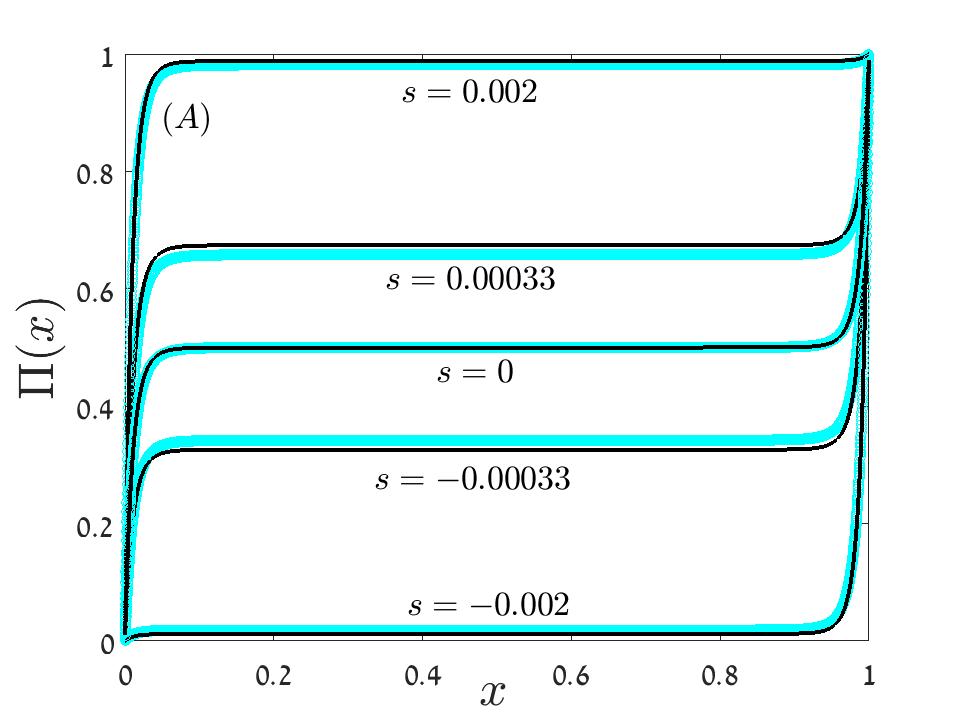}
\includegraphics[width=8cm]{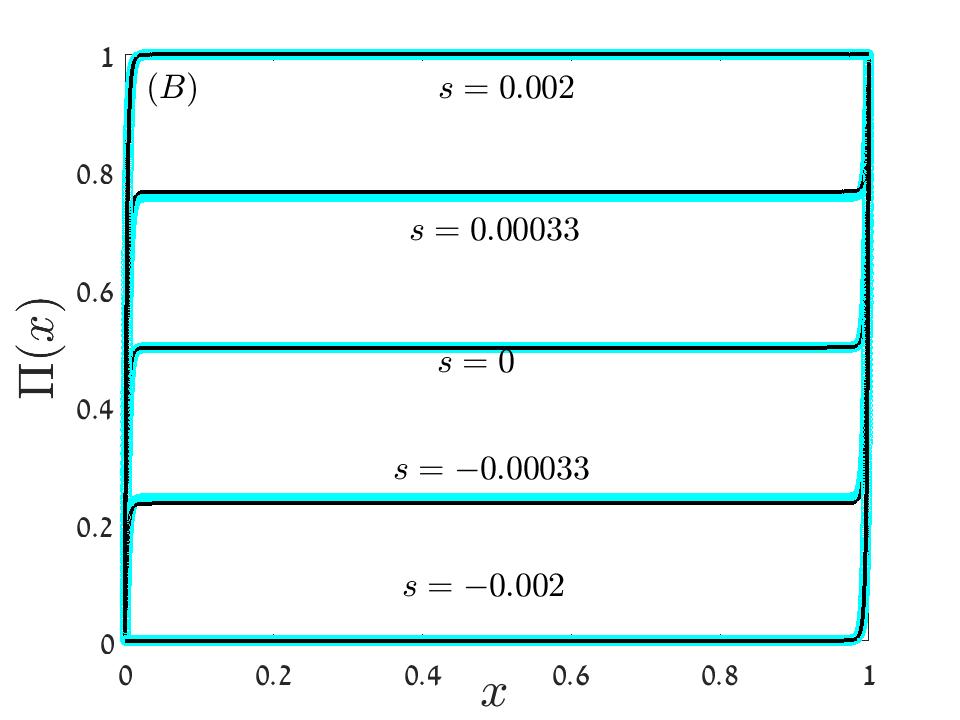}
\caption{$\Pi(x)$ vs. $x$ for model B. In both figures $\gamma = 0.2$ and $\delta = 0.1$, in panel (A) $N=5000$ and in panel (B) $N=20,000$. Numerical solutions of the discrete equation (\ref{eq60a}) (blue circles) are compared with the uniform approximations (\ref{unifB}) (full black lines) for $s_0=0$ ($\tilde{s} =0, \ n_c = \infty$), $s_0= \pm 0.00033$ ($|\tilde{s}| =0.33, \ n_c \sim 10^4$) and $s_0= \pm 0.002$ ($|\tilde{s}| = 10$). The pronounced plateau in which  $\Pi(x) =C_1$, where $C_1$ is neither zero nor one,  exists when $|\tilde{s}|<1$. As  $N$ growth  the value of $C_1$ increases (for positive $s_0$) or decreases (for negative $s_0$), as one may notice  by comparing the lines for $s_0= \pm 0.00033$ in the two panels.       }\label{fig2}
\end{figure}

The main difference between Eqs. (\ref{modelB1}) and (\ref{modelB2}) and their model A counterparts, Eqs. (\ref{modelA1}) and (\ref{modelA2}), is the different scaling of  $C_2$. In model B, $C_2$ goes to zero when $N \gg \exp(\delta)/g$. Above this $s_0$-independent point the chance of fixation in the middle region is fixed, $C_3=C_1$, as demonstrated in Figure \ref{fig2}. A wide plateau appears in the middle region due to the force towards $x^*$ which is so strong that $\Pi$ becomes almost $x$-independent.

The uniform solution when $C_2 \to 0$ has a relatively simple form (see Figure \ref{fig2}),
\begin{equation} \label{unifB}
\Pi(x) = C_1 [1-(1+Gx)^{-\alpha -1/\delta} - [1+G(1-x)]^{\alpha -1/\delta}]+[1+G(1-x)]^{\alpha -1/\delta},
\end{equation}

\begin{figure}
\includegraphics[width=8cm]{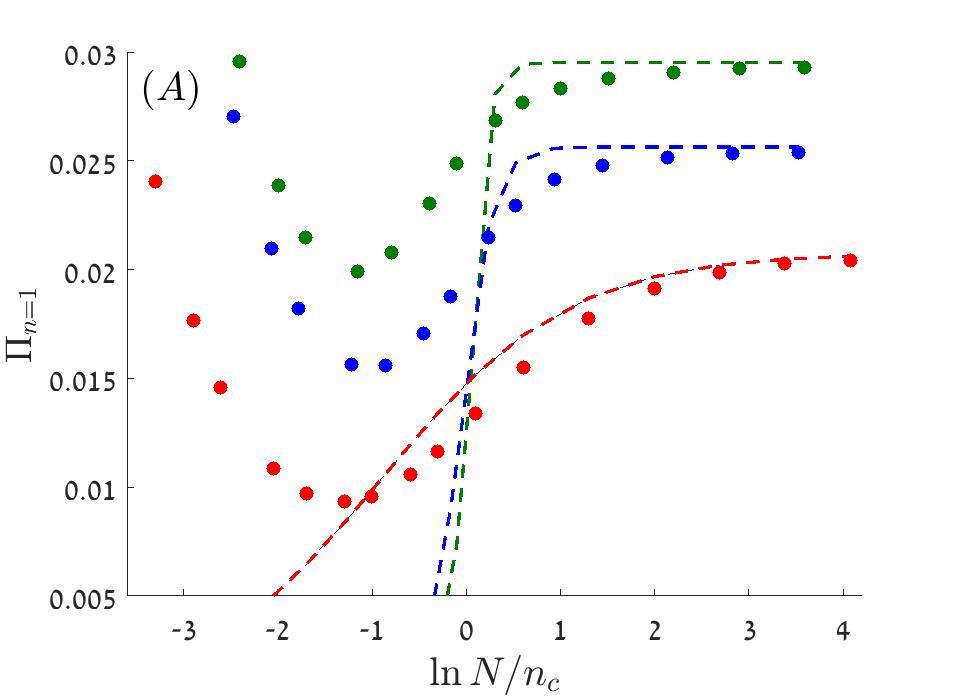}
\includegraphics[width=8cm]{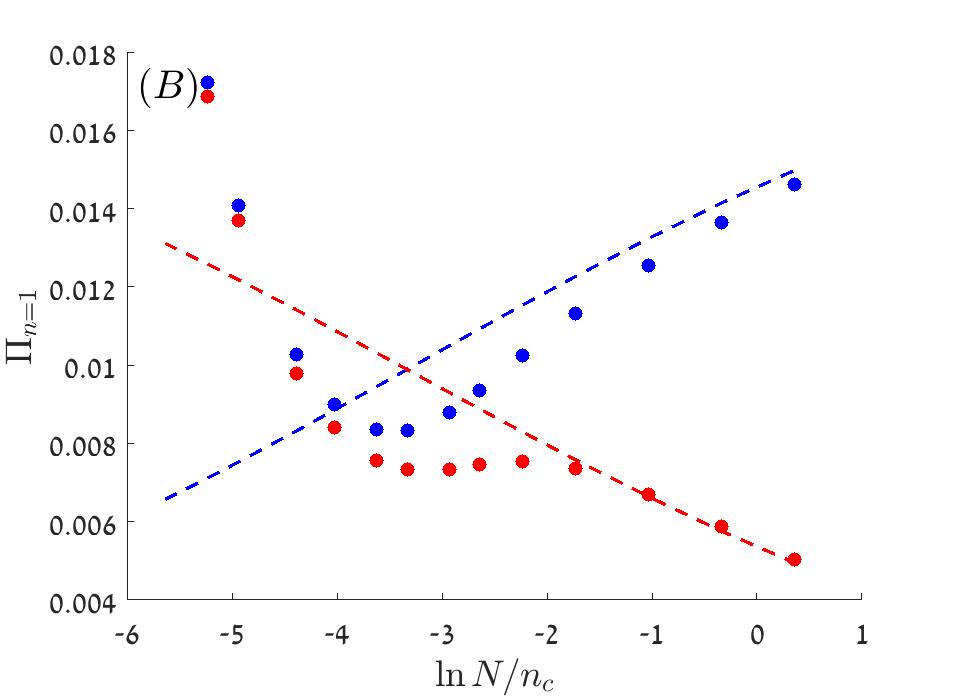}
\caption{The  chance of fixation for the lineage of a single mutant, $\Pi_{n=1}$, is plotted against the effective community size $N/n_c$ on a semi-logarithmic scale (the $y$ axis is linear, as opposed to Fig. \ref{fig3}). Parameters are $\gamma = 0.2$ and $\delta = 0.1$, so $G = Ng = 1$ corresponds to $N=500$, which is the seventh point in each dataset. Filled circles represent the results of a numerical simulation and the dashed lines are the prediction of Eq. (\ref{smB}). The actual values of $N$ used here are between $10$ to $20,000$ (for $s_0=0.001$, $N$ goes up to $80,000$).  In panel A the results are shown for  $s_0=0.001$ (red) $s_0=0.06$ (blue) and $s_0=0.01$ (green). The chance of fixation \emph{grows} with $N$ and becomes $N$ independent in the strong selection regime. Panel B shows the results for the weak selection ($N \ll n_c$) regime for both positive and negative selection, $s_0 = 0.0003$ (blue) and  $s_0 = -0.0003$  (red).  }\label{fig7}
\end{figure}

 and the chance of fixation of a single mutant ($x = 1/N$) is,
 \begin{equation} \label{smB}
 \Pi_{n=1} = \frac{1-\frac{1}{(1+g)^{\frac{1}{\delta}(\tilde{s}+1)}}}{1+D_1 G^{-2 \alpha}}.
 \end{equation}

 Amazingly, $\Pi_{n=1}$ turns out to be an \emph{increasing} function of $N$, a behaviour that  manifests itself in Figure \ref{fig7}. This phenomenon reflects the stabilizing effect of the nonlinear mechanism: the chance to reach the plateau does not depend on $N$ because the plateau occurs at values of $x$ that scale like $1/N$.  For example, in the large $N$ limit $\Pi_{s_0 = 0}(x)$ sticks to $1/2$ for any $x(1-x) > 2/(N\gamma^2)$. For $G \ll 1$ the chance of fixation  decays like $1/N$ since this regime (in which our expressions fail) is dominated by demographic stochasticity. When $N$ increases  the stabilizing mechanism wins against the demographic noise and leads to an increase of the chance of fixation.

 $\Pi_{n=1}$ in Eq. (\ref{smB}) is a multiplication of two factors:  its numerator is the chance of establishment $\Pi_{est}$, which is the probability that the mutant population will reach the basin of attraction of the coexistence fixed point (the plateau).  $C_1$, that determines the denominator,  is the chance that the mutant population will reach fixation given  establishment. Accordingly, for a single  beneficial mutant,
\begin{equation}
\Pi_{est} = 1-(1+g)^{-\alpha-1/\delta}.
\end{equation}

 If the mutant is advantageous ($s_0 > 0$) and the system is in its  strong selection regime ($G^{-2\alpha} \to 0$ or  $N \gg n_c$), $C_1 = 1$ and $\Pi_{n=1} \approx \Pi_{est}$. In terms of $\tilde{s}$,
 \begin{equation}
 \Pi_{n=1} = 1-\frac{1}{(1+g)^{\frac{1}{\delta}(\tilde{s}+1)}}.
 \end{equation}
  This is a monotonously decreasing function of $\delta$. When $\delta$ increases, the stochasticity becomes stronger and the stabilizing mechanism weakens~\cite{danino2016effect}, both effects tend to decrease the chance of fixation. The dependence on  the amplitude of environmental fluctuations, $\gamma$,  is more complicated. When $\gamma$ is small, its increase facilitates the stabilizing mechanism that increases the chance of fixation, while for large $\gamma$ the increase in $n_c$ is the dominant effect and $\Pi_{n=1}$ decreases.

 Model A and model B differ even more dramatically in the weak selection regime $N \ll n_c$, where $G^{-\alpha} \approx 1-\alpha \ln G $. For model B, the chance of fixation becomes,
 \begin{equation} \label{fB}
 \Pi_{n=1} \approx \frac{1-(1+g)^{-\frac{1}{\delta}(\tilde{s}+1)}}{2}\left(1+\frac{\tilde{s} \ln(Ng)}{\delta}\right).
 \end{equation}
 Unlike model A, where the chance of fixation decays logarithmically with $N$ in the weak selection regime, here the denominator of (\ref{fB}) is an ${\cal O}(1)$ constant.  When  $\tilde{s} \ll 1$, for example, the chance of fixation is one half of the chance of establishment: the effective strength of the selection bias is zero, so once the mutant population reaches the plateau its odds to win or to lose are equal. For nonzero $\tilde{s}$ there is a linear increase or decrease of $\Pi_{n=1}$ as a function of $\ln N$. This relatively weak effect is demonstrated in panel (B) of Fig. \ref{fig7}.

Before concluding this section we would like to add a technical comment.  $\Pi(x)$ is obtained from Eq. (\ref{eq60a}) by solving a linear problem (dividing a matrix by a vector). Using the sparsity of the matrix we were able to analyze systems with up to $N=10^6$ individuals. Because of the plateau that characterizes model B in the strong selection regime, this numerical solution becomes difficult; the plateau indicates that the matrix to be inverted is almost singular. To solve that we have used quadrapole precision and this makes the numerics much slower and limits  available system sizes to $N$ values up to $20,000$.

\section{The chance of fixation for a deleterious mutant  under strong selection - A WKB approach} \label{tunnel}

 Until now we discussed the weak selection regime for both beneficial ($s_0 >0$) and deleterious ($s_0 <0$) mutants, but the strong selection regime ($N \gg n_c$) for deleterious mutant  has not yet  been considered.  In this section we would like to provide a few basic insights for that case.

 Quantitatively, one may guess that the chance of fixation in this regime behaves differently  when $\gamma < |s_0|$ and $\gamma>|s_0|$. In the latter case, the growth rate of a deleterious mutant is still positive ($\gamma - |s_0|$) during half of the time, so the most probable (yet rare) route to fixation is based on picking a  sequence of good years. During this series of lucky events the (on average) deleterious mutant plays the role of a beneficial one, and its time to fixation scales like $\ln N$. Therefore, the chance of fixation, namely the chance to pick such a lucky sequence, decays like a power-law in $N$.  On the other hand, when $\gamma < |s_0|$,  the route to fixation involves an improbable series of successes in consecutive elementary duels (reflecting demographic stochasticity) and in such a  case $\Pi_{n=1}$ decays exponentially in $N$, like in the purely demographic case (\ref{eq1}).

 Looking at the equations above, one notices that the decay of $\Pi_{n=1}$ when $s_0<0$ (and $\alpha <0$) is due to the divergence of the $G^{-2\alpha} = G^{2|s_0|/g}$ term in  the denominator of $C_1$ when the selection is strong. Accordingly, our theory predicts in that regime (as suggested in \cite{assaf2013cooperation} for model A with $G \gg 1$) a power-law decay, $\Pi_{n=1} \sim N^{-4|s_0|/\gamma^2 \delta}$. The exponent of $N$ grows with $s_0$ and shrinks when the environmental stochasticity become stronger, as expected. However it does not show any qualitative shift when $\gamma = |s_0|$.

 This difficulty turned out to be related to the failure of the continuum approximation that has been used when we have translated the difference equations (\ref{eq60a}) to  the differential equation (\ref{eq10}). As explained in \cite{kessler2007extinction}, this procedure fails when the differences between neighboring points (say, $\Pi_{n+1} -\Pi_n$) are too large and cannot be approximated using first and second derivatives. To overcome this obstacle a WKB approach was suggested~\cite{kessler2007extinction}; here we would like to implement it for a model with environmental stochasticity. We shall neglect, for the moment, the effect of demographic noise and assume that extinction and fixation happen when the abundance reaches $1/N$ and $1-1/N$, correspondingly.

As explained in the introduction [following Eq. (\ref{logistic})], in the absence of demographic noise and under model A dynamics $\dot{z} = sz$, where $z \equiv x/(1-x)$.  Accordingly, during $\delta$ generations the dynamics of $z$ satisfies,
 \begin{equation}
 z(t+\delta) = z(t) e^{ - (|s_0| \pm \gamma)\delta},
 \end{equation}
 so one may consider the stochastic process as a biased random walk along the $\ln(z)$ axis. The random walker picks a left or a right move with equal chance $1/2$, but left moves towards extinction ($\ln(z) \to \ln(z) -(|s_0| + \gamma)\delta \equiv \ln(z) - \ell_1$) are longer than right moves towards fixation ($\ln(z) \to \ln(z)+ (\gamma-|s_0|)\delta \equiv \ln(z) + \ell_2$). The backward Kolmogorov equation is,
 \begin{equation} \label{WKB1}
 \Pi(\ln z) = \frac{1}{2} \Pi (\ln z-\ell_1) + \frac{1}{2} \Pi(\ln z +\ell_2) ,
 \end{equation}
  with the boundary conditions  $\Pi(\ln z=-\ln N)=0$ and $\Pi(\ln z=\ln N)=1 $.

\begin{figure}
\includegraphics[width=8cm]{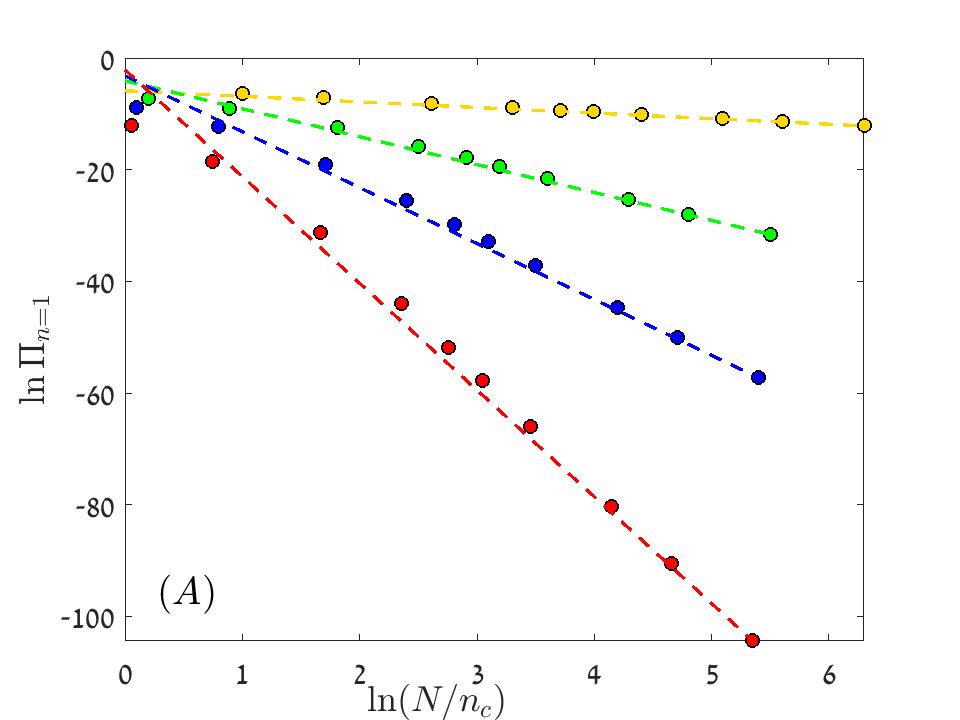}
\includegraphics[width=8cm]{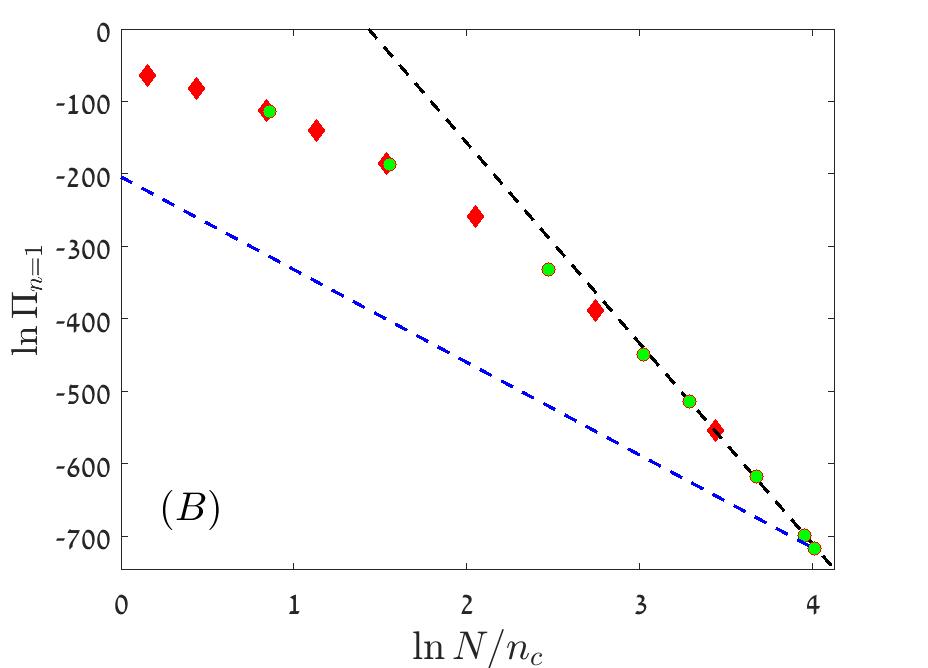}
\caption{$\ln \Pi_{n=1}$ vs. $\ln N/n_c$ ($n_c$ is defined with the absolute value of $s$) for a deleterious mutant in the strong selection regime ($G^{2|s_0|/g} \gg 1$). Panel (A) shows results for model A  in the small $Q'$ regime.  Parameters are $\gamma =0.2$, $\delta = 0.1$ and $(-s_0)$ takes the values  $0.001$ (yellow), $0.005$ (green), $0.01$ (blue) and $0.019$ (red). Filled circles are the results obtained from the numerical solution of the discrete equations (\ref{eq60a}) and the dashed lines have the slope $-4s_0/\gamma^2 \delta$. Similar results were obtained for model B. In  panel  (B) the power of our WKB technique is demonstrated.  Here $\gamma = 0.25$, $\delta =0.1$ and $s_0 = -0.2$, model A results are presented as green circles while model B results are red diamonds.  The slope suggested by  the small $Q'$ approximation (blue dashed line, with slope $-4s_0/(\gamma^2 \delta)=-128$) clearly fails to describe the large $N$ behavior. A much better fit is provided by the black dashed line, with a slope $-2Q' = -277$  that was obtained from a numerical solution of Eq. (\ref{WKB3}).  The intercepts of the dashed lines in both panels were chosen manually  such that each line fits the last point of the corresponding dataset. }\label{fig5}
\end{figure}

 To implement the WKB technique, one writes $\Pi(\ln z) = \exp(Q[\ln z])$ and $\Pi_{\ln z  \pm \ell} = \exp(Q[\ln z \pm \ell]) \approx \exp(Q[\ln z]) \pm \ell Q')$, where $Q' = \partial Q(\ln z)/\partial \ln z$. In this WKB  formalism we implement  the continuum approximation to $Q$, i.e., for the logarithm of $\Pi$. Eq. (\ref{WKB1}) then takes the form,
 \begin{equation}\label{WKB2}
 1 =  \frac{1}{2} \left(e^{-\ell_1 Q'} + e^{\ell_2 Q'} \right).
 \end{equation}
 and yields the transcendental equation,
  \begin{equation}\label{WKB3}
 \cosh(\gamma \delta Q') = e^{s \delta Q'}.
 \end{equation}
 Since $Q'$ is $\ln(z)$ independent, $\Pi \sim exp(Q' \ln(z))$, and given the boundary conditions one obtains,
 \begin{equation}
 \Pi(\ln z) = \frac{e^{Q' \ln(z_0)} - e^{-Q' \ln N}}{e^{Q' \ln(N)} - e^{-Q' \ln N}}.
 \end{equation}
  If  a "single mutant"  is associated with $x_0=2/N$  (since we impose the boundary condition at $1/N$, we have to define it that way, but the results must be independent of this choice)  the chance of fixation decays like a power-law in $N$,
 \begin{equation}
 \Pi_{n=2} \sim e^{-2 Q' \ln(N)} = N^{-2Q'}.
 \end{equation}
 The value of $Q'$ is given by  (\ref{WKB3}). For small $Q'$, $$Q' \sim 2s_0/\delta \gamma^2 = s_0/g,$$ in agreement with the definition of $C_1$ above.  On the other hand, if  $ Q'$  is large,
  \begin{equation}
  Q' \sim \frac{\ln 2}{\delta(\gamma -s_0)},
  \end{equation}
  and this expression diverges when $\gamma \to |s_0|$, as required, to mark the transition to the exponential phase. Between these two limits, the expression $$ Q' \sim \frac{2s_0}{\delta(\gamma^2-s_0^2)},$$ provides a decent approximation. The accuracy of this WKB argument is demonstrated in Figure \ref{fig5}.

This WKB argument gives the decay of $\Pi$ in the power-law regime. When $\gamma \ge |s_0|$ it fails, of course, since stochastic tunneling in this regime  occurs only due to demographic stochasticity that was neglected in (\ref{WKB1}). For further discussion of the exponential phase (in the context of extinction times) see \cite{ovaskainen2010stochastic,kamenev2008colored}.

\section{Discussion}

The two models considered in this paper have one important feature in common: the abundance scale $n_c = \exp[g/(2|s_0|)]/g$, below which the mutant population dynamics is dominated by fluctuations and above it by selection. This scale may become extremely large when the differences in the mean fitness are much smaller that the amplitude of the  temporal fitness fluctuations, and one may easily imagine a situation where it becomes comparable or even larger than the effective size of an empirical community, meaning that the ecological or the evolutionary process takes place in the weak selection regime.

In the opposite, strong selection phase, the qualitative features of the chance of fixation $\Pi_{n=1}$ are not much different between model A and model B. In both cases the chance of fixation for a beneficial mutant is $N$-independent while the chance of fixation of a deleterious mutant falls like $N^{-2Q'}$ when $\gamma > |s_0|$ and exponentially with $N$ if $\gamma <|s_0|$. In model A the chance of fixation in the strong selection regime decreases as the environmental noise becomes stronger, while in model B  it decreases with the correlation time $\delta$ but increases with noise amplitude $\gamma$. (Through this discussion, when the features of  model B are contrasted with those of model A,  model B  is assumed to support a noise induced attractive fixed point, i.e., $|\tilde{s}|<1$. Otherwise, the behavior of model B dynamics is qualitatively the same as model A).

On the other hand, in the weak selection regime there are substential differences between the two scenarios. As required by its name, in this regime selection is a second order effect and the fate of the mutant population is determined by stochasticity. In a stochastic and balanced game, like the classical  gambler's ruin problem, the chance to win is inversely proportional to the effective size of the community,  so under purely demographic noise it is $1/N$ and under model A dynamics $1/\ln N$.

In sharp contrast with this result, in model B the system supports an attractive fixed point at $x^*$. The plateau that characterizes $\Pi(x)$ in that case (Figure \ref{fig2}) reflects the effect of this attractive fixed point, marking the range of $x$ values which lies in its basin of attraction. The attractiveness of this coexistence point grows with $N$ and leads to an apparently paradoxical behavior: an increase of the chance of fixation with $N$.    Once this fixed point becomes dominant, the fate of the mutant population depends on its chance of establishment $\Pi_{est}$, i.e., of reaching the plateau, and  on $C_1$,  the probability to  jump from the plateau region to fixation. Since the plateau is wide, these two probabilities are $N$ independent, and so is the chance of fixation itself. Even if $N$ is huge, as long as it is much smaller than $n_c$, model B yields an $N$-independent value for $\Pi_{n=1}$ for both beneficial and deleterious mutants.

Even in the strong selection regime, where the chance of fixation of a deleterious mutant are vanishingly small, model B dynamics still supports an attractive fixed point at $x^*$ as long as $|\tilde{s}|<1$. The chance of establishment is still $N$-independent; it is the chance of fixation conditioned on establishment,  $C_1$, which goes to zero in that case. As shown in~\cite{danino2016stability}, the lifetime of the mutant population, once established, is $N^{(1-|\tilde{s}|)/\delta}=N^{1/\delta-\alpha}$ generations, a huge time for large communities. The stabilizing mechanism of model B thus allows for the  long-term persistence of a macroscopic, ${\cal O}(N)$, extinction-prone population with negative fitness.

This phenomenon may provide a plausible mechanistic explanation to one of the the mysteries of evolutionary dynamics: the ability of evolutionary pathways to cross fitness valleys, i.e., to sustain a chain of suboptimal intermediate forms that bridge between two fitness peaks in a rugged fitness landscape. This stochastic tunneling has been recognized a while ago as a major theoretical problem, since the chances for a tunneling event are vanishingly small \cite{gavrilets2010high}. To overcome this problem modern theoretical studies consider evolution on a neutral or nearly neutral (holely, high-dimensional, connected) fitness landscape. In this neutral picture  a separate mechanism has to be invoked to explain speciation, while on rugged landscape each species corresponds to a separate fitness pick and disruptive selection is guaranteed by the landscape itself. The long term existence of macroscopic suboptimal populations, considered here in  context of model B, may allow for such a tunneling to occur with relatively high probability through a chain of mutation as long as the depth of the fitness valley is smaller than $\gamma^2/2$, while keeping the intermediate forms extinction-prone.

The relevance of the mechanisms suggested here to the development of natural communities depends on the amplitude of fitness variations with respect to their time-averaged value, on the typical correlation time of these fluctuations and  on the range of competition - whether it is local/pairwise (model A) or global (model B). It is quite difficult to quantify $s_0$ and $\gamma$ from field data, and in experimental systems the external conditions are usually kept fixed, as opposed to the intrinsic variability of natural environments. Still, we believe that the theory presented here, when applied to some experiments and to field data in population genetics and community ecology, may suggest many new insights into the processes that govern the composition and the evolution of natural communities.

\section{\bf Acknowledgments} This research  was supported by the ISF-NRF Singapore joint research program (grant number 2669/17).

\bibliography{refs_immanuel}

\clearpage

\end{document}